\documentclass[12pt,a4paper]{article}

\begin{document}

\newcommand*\Z{\mathrm{Z}\hspace{-2mm}\mathrm{Z}}
\newcommand*\C{\mathrm{l}\hspace{-2mm}\mathrm{C}}
\newcommand*\R{\mathrm{I}\!\mathrm{R}}
\newcommand*\N{\mathrm{I}\!\mathrm{N}}
\newcommand*\T{\mathrm{T}}

\newcommand*\st{\mathrm{st}}
\newcommand*\tg{\mathrm{tg}}
\newcommand*\NS{\mathrm{NS}}
\newcommand*\fin{\mathrm{fin}}
\newcommand*\ONB{\mathrm{ONB}}
\newcommand*\LF{\mathrm{LF}}
\newcommand*\LFN{\mathrm{LFN}}
\newcommand*\Pol{\mathrm{Pol}}

\title{An approach to nonstandard quantum mechanics}

\author{A. Raab\\ 
St.-Ilgener-Str. 2, 69190 Walldorf \\
Federal Republic of Germany\\
E-mail: andreas.raab.mail@web.de}

\date{\today}

\maketitle

\begin{abstract}
We use nonstandard analysis to formulate quantum mechanics in hyperfinite-dimensional spaces. 
Self-adjoint
operators on hyperfinite-dimensional spaces have complete eigensets, and bound states and 
continuum states of a Hamiltonian can thus be treated on an equal footing. 
We show that the formalism extends the standard formulation of quantum mechanics. To this end
we develop the Loeb-function calculus in nonstandard hulls. The idea is
to perform calculations in a hyperfinite-dimensional space, but to interpret expectation 
values in the corresponding nonstandard hull. We further apply the framework to
non-relativistic quantum scattering theory. For time-dependent scattering theory, we 
identify the starting time and the finishing time of a scattering experiment, and we obtain
a natural separation of time scales on which the preparation process, the interaction
process, and the detection process take place. For time-independent scattering theory, we 
derive rigorously explicit formulas for the M\o ller wave operators and the S-Matrix.
\end{abstract}

\section{Introduction}
\label{INTRO}

Quantum mechanics is conventionally formulated in a complex Hilbert space, ${\cal H}$. 
The possible states of a quantum system are associated with unit vectors in ${\cal H}$, 
and the observables are associated with self-adjoint linear operators on ${\cal H}$.
A central role plays the Hamiltonian of the quantum system. The eigenvalues of the Hamiltonian
are commonly interpreted as the energies of bound states of the system. Moreover, the values of 
the continuous part of the Hamiltonians spectrum are interpreted as the energies of 
''continuum'' or scattering states. The interpretation is physically motivated, and we use the
term ''continuum state'' in a physical sense in the following. 
While bound states can be identified with eigenvectors of the 
Hamiltonian, appropriate vectors do not exist in ${\cal H}$ for continuum states. 
Continuum states can be treated in the conventional Hilbert-space framework only approximately
\cite{ree72}.

The probably most prominent approach to solve this problem is the 
rigged-Hilbert-space formalism \cite{ant_boh98,boh_boh98}. 
Within the rigged Hilbert-space formalism,
continuum states are associated with linear functionals on a dense subspace ${\cal M}$ of 
${\cal H}$. The linear functionals belong to the dual space ${\cal M'}$ of ${\cal M}$, which 
is a locally convex space and not a Hilbert space.
In particular, we do not have a scalar product at hand. 
The rigged-Hilbert-space formalism is thus 
more complicated than the Hilbert-space formalism, and more efforts are required in 
mathematically rigorous applications. 
There exist further similar approaches to model continuum states, e.g. 
approaches that introduce lattices of Hilbert spaces or partial-inner-product spaces. 
These two approaches together with the rigged-Hilbert-space approach 
can be classified as ''super-Hilbert-space'' formalisms \cite{ant_boh98}.
The main idea in common is to use linear functionals on (dense) 
subspaces of ${\cal H}$, and the mathematical complications are consequently of the same nature.
If we look for an appropriate formalism, which includes a scalar 
product, we still face a mathematical-modelling problem.

This article suggests an approach to nonstandard quantum mechanics. This means that we use
nonstandard analysis (NSA) to construct a framework where quantum mechanics can be formulated
without the drawbacks mentioned above. As far as I know, the first work in this direction 
is done in \cite{far75:177}. However, this article presents another approach that 
focuses more on eigenvector expansions. Eigenvector expansions are of great 
importance in practical applications since they simplify calculations.
Within the approach, we are able to treat bound and continuum
states on the same footing without losing the scalar product. However, we require a 
basic knowledge of NSA for the construction as it is presented in \cite{nsa97}, for example. 
Nevertheless, I make a few remarks on NSA in the following.

NSA has its origin in logics and provides an astonishing rich formalism. I believe that 
we should rather speak of nonstandard methods than of NSA since NSA suggests an application of 
nonstandard methods to analysis, but the term NSA is commonly used in the literature in a 
general sense. In this sense, NSA can be applied to a wide range of mathematical areas, e.g. 
real analysis, topology, measure theory, functional analysis, etc.. NSA introduces rigorously 
many interesting objects like infinitesimals, infinitely large numbers, and functions that 
behave like Dirac's delta distribution. I believe that these objects - which are not available
in standard mathematics - make NSA rather attractive for physical applications.

The basic tool of NSA is the transfer principle. The transfer principle enables us to transfer
sets and formulas from standard to nonstandard frameworks. Roughly
speaking, every formula in a standard framework is true if, and only if, the corresponding 
transferred formula is true in the nonstandard framework. For the application of the 
transfer principle we however have to formulate formulas in a rather 
formal way marked by logics. The formal language appears laborious from the viewpoint of the 
concrete application, but we need the language to ensure rigorous results.

The article is organized as follows. In Sec. \ref{BF}, the basic framework is introduced, 
and we start discussing its relationship to the standard framework used in
quantum mechanics. We continue the discussion in Sec. \ref{FC} where we derive the 
required nonstandard function calculus. We establish then the final form of 
the approach to nonstandard quantum mechanics in Sec. \ref{QM}. In Sec. \ref{ST}, we 
apply the framework to non-relativistic scattering theory. We discuss first the impact
of the approach on time-dependent scattering theory, and derive then explicit
formulas within time-independent scattering theory. Finally, we summarize the results
and conclude in Sec. \ref{Conc}.

\section{Nonstandard extensions}
\label{BF}

\subsection{Spatial and operational extensions}
\label{SOEX}

NSA basically introduces extensions of superstructures that contain for a given context
all mathematical objects of interest. Superstructures usually contain real numbers, complex
numbers, functions, etc.. The extension of a superstructure ${\cal V}$ is actually an 
injective mapping of ${\cal V}$ onto another superstructure ${\cal W}$. 
We note that there exist different types of extensions, but for applications of NSA 
polysaturated extensions are probably most convenient. For this reason, we assume a 
polysaturated extension $^\star: {\cal V} \to {\cal W}$ of a superstructure
${\cal V}$ that contains every standard mathematical object we will consider in this article.
Roughly speaking, the extension allows us to switch from the 'standard world' ${\cal V}$ to the 
'nonstandard world' ${\cal W}$, in which we can use our standard mathematical objects of interest 
more conveniently, as we will see. In particular, ${\cal V}$ contains a complex Hilbert space ${\cal H}$ 
and linear operators on ${\cal H}$, which implies that appropriate counterparts are available in the
'nonstandard world' ${\cal W}$. Moreover, for the sake of convenience we follow the common
practice and drop the prefix $^\star$ from many nonstandard objects when there is
no ambiguity. For example, we write  $r + s$ instead of $r \,\, ^\star \!\!+ s$ for $r,s \in 
\,^\star\R$. We obtain for our example a sensible simplification of 
notation since two hyperreal numbers can be added in the same manner as two real numbers.

As a consequence of polysaturation, there exists a 
hyperfinite-dimensional space $H$ which externally contains ${\cal H}$ \cite{wolff_nsa97}, i.e.
$\{ ^\star x \, : \, x \in {\cal H} \} \subset H$. For the sake of convenience we simply
write ${\cal H} \subset H \subset \, ^\star {\cal H}$. We use this result as
an avenue to a nonstandard framework for quantum mechanics. Beside the well-known advantages
of NSA, which are sketched in Sec. \ref{INTRO}, this approach has another important advantage:
The transfer principle allows us to apply the results of 
linear algebra in finite-dimensional spaces to the hyperfinite-dimensional space $H$.
We note that this idea is already known in the literature for a long time 
\cite{ber66:421, ber72:419}.

The main goal in the following is to show that the formulation of quantum mechanics in
an appropriate hyperfinite-dimensional space yields an extension of the 
standard formulation of quantum mechanics. We choose however a more general setting than
outlined above, and assume only a dense subspace ${\cal M}$ of ${\cal H}$. We will see in Sec. 
\ref{SO} that this approach is quite convenient when we construct the hyperfinite dimensional
space $H$ in an example. However, by polysaturation there exists a hyperfinite-dimensional 
space $H$ which externally contains ${\cal M}$, i.e. ${\cal M} \subset H \subset \, 
^\star {\cal M} \subset \, ^\star {\cal H}$.\\[.3cm]
{\bf Proposition 1:} Let $H \subset \, ^\star {\cal H}$ be an internal Hilbert space 
that externally contains a dense subspace ${\cal M}$ of ${\cal H}$, 
and let $O$ denote the projection of $^\star{\cal H}$
onto $H$, then $^\star x \approx O\, ^\star x$ for all $x \in {\cal H}$.
\\[.15cm]
{\em Proof}: Assume $x \in {\cal H}$. Since ${\cal M}$ is dense in ${\cal H}$ the
internal statement $(\exists y \in H) \| ^\star x - y \| < 1/n$ is true for all $n \in \N$.
By the overflow principle there exists an infinite $m \in \, ^\star \N$ for which
$(\exists y \in H) \| ^\star x - y \| < 1/m$ holds. Moreover, the (transferred) projection theorem
states that $\| ^\star x - O \, ^\star x \| \leq \| ^\star x - y \| < 1/m$ \cite{ree72}.
$\diamondsuit$ \\*[.3cm]
Let us proceed with our discussion. As in proposition 1, let $O$ denote the projection 
of $^\star {\cal H}$ onto $H$. $O$ can be defined (as usual) by an orthonormal basis of $H$,
as we show in the appendix. Moreover, let $A$ be a self-adjoint operator with domain 
${\cal D} (A)$, for which ${\cal M} \subset {\cal D} (A) \subset {\cal H}$ holds. 
Using the transfer principle
we define the restriction of $^\star A$ to $H$ by $B = O \, ^\star A O$, and 
$B \, ^\star x = O \, ^\star (A x) \approx \, ^\star (A x)$ for all 
$x \in {\cal M}$. $B$ can thus be seen as a nonstandard extension of the restriction of $A$
to ${\cal M}$. As shown in the appendix, $B$ is an internal
hyperfinite-rank operator, and there exists an eigensystem 
$\{ (\lambda_i, x_i) \}_{i=1}^h $ for $B$, which yields the representation
\[
(\forall x,y \in H) \quad \langle x, B y \rangle =
\sum_{i=1}^h \lambda_i \, \langle x, x_i \rangle \, \langle x_i, y \rangle .
\]
We note that $h$ is the (nonstandard) dimension of $H$.

Our construction shows that there exist self-adjoint hyperfinite-rank operators, which 
yield extensions of standard self-adjoint operators in a certain sense. The relation is however
rather weak at the moment since we do not know how the spectra and the function calculus
of these operators are related. In particular, we may ask if 
nonstandard eigenvalues and nonstandard eigenvectors can be interpreted in a sensible way.
To discuss these questions more thoroughly 
we introduce nonstandard hulls, which are important tools in NSA \cite{wolff_nsa97}. 
From a physical point of view,
the introduction of nonstandard hulls is motivated by the assumption that infinitesimally
different states cannot be distinguished in a measurement. 

The definition of the nonstandard hull $^o H$ of the hyperfinite-dimensional space 
$H$ introduces an equivalence relation on
the set of finite nonstandard vectors, $ \fin( H ) = \{ x \in H : \, 
\| x \| \in \fin( ^\star \R ) \}$. We note that $\fin( ^\star \R )$ is the set of finite
hyperreals. Two vectors are equivalent if their difference has
infinitesimal norm. The nonstandard hull $^oH$ of the 
space $H$ is given as the quotient
\begin{eqnarray}
^o H & = & \fin( H ) / H_0 , \quad  H_0 = \{ x \in H : \, 
\| x \| \approx 0 \} ,\\
\nonumber
\| ^o x \| & = & \st( \| x \| ) , \quad 
\langle ^o x , \, ^o y \rangle = \st( \langle x, y \rangle ) .
\end{eqnarray}
$^o H$ is a Hilbert space, and ${\cal H}$ is a closed subspace of $^o H$ by proposition 1. 
Especially, reconsidering our example above the equation 
$B \, ^\star x = O \, ^\star (A x) \approx \, ^\star (A x)$ yields
$Ax = \, ^o(B x)$ for all $x \in {\cal M}$.
We note that we adopt the notation $^o H$ for the nonstandard hull of $H$ as it is used in
\cite{ber66:421, ber72:419}, instead of the notation $\hat{H}$ that seems to be more common
\cite{nsa97}. I believe that we gain a more uniform
notation since, for example, for the standard part $^o r = \st (r)$ of $r \in \fin (
^\star \R)$ we could also write $\hat{r}$, 
meaning an element of the nonstandard hull of $^\star \R$. 

Furthermore, nonstandard hulls are defined also for finitely bounded nonstandard linear operators. 
If $A$ is a bounded operator on ${\cal H}$, then $B = O \, ^\star A O$ is finitely bounded and
its nonstandard hull is defined as
\begin{eqnarray}
^o B  \, ^o x & = & ^o (Bx) \quad \forall \, x \in \fin(H) \\
\nonumber
\| ^o B \| & = & \st ( \| B \| ) .
\end{eqnarray}
Since $A$ is bounded we obtain $^\star(Ax) \approx O \, ^\star A O \, ^\star x$ for $x \in
{\cal H}$. Thus,
$^o B$ is a self-adjoint bounded operator on $^o H$, which extends $A$, i.e.
$^o B x = \, ^o( B \, ^\star x ) = A x$ for all $x \in {\cal H}$. We note that operational
nonstandard extensions of this type are already discussed in \cite{ber72:419}. In particular,
the relationship between the spectral resolution of bounded self-adjoint operators and their
nonstandard extension is investigated. However, we come back to these results in Sec. \ref{SPEX}.

\subsection{Spectral properties of operator extensions}
\label{SPEX}

Hyperfinite-rank operators have convenient spectral properties, which 
are determined by linear algebra. If we consider a nonstandard extension $B$ of 
a self-adjoint operator $A$ as constructed in Sec. \ref{SOEX} then we may naturally ask
how the eigenvalues and eigenvectors of the hyperfinite-rank operator $B$ are related to 
the spectral resolution of $A$. Let us focus first on the eigenvalues.
\\[.3cm]
{\bf Lemma 1:} Let $B$ be a normal hyperfinite-rank operator, and let
$\lambda \in \C$. Assume that for each $n \in \N$ there
exists an $x_n \in H$ for which $\| x_n \| \approx 1$ and $\| \lambda x_n - B x_n \| <
1/n$ holds, then there exists a $\lambda' \in \sigma(B)$, and $\lambda \approx 
\lambda'$.
\\[.15cm]
{\em Proof}: As shown in the appendix, $B$ has an 
eigensystem $\{ (\lambda_i, x_i) \}_{i=1}^h $. 
Fix $\epsilon \in \R_+$, $\epsilon < 1$, then
\[
(\forall n \in \N) (\exists x \in H ) \, \left( \| x \| > (1- \epsilon) \, \wedge 
\, \| \lambda x - B x \| \leq \frac{1}{n} \right).
\]
By the overspill principle there exists an infinite $m \in \, ^\star \N$
for which 
\[
(\exists x_0 \in H ) \, \left( \| x_0 \| > (1- \epsilon) \, \wedge 
\, \| \lambda x_0 - B x_0 \| \leq \frac{1}{m} \right)
\]
is true. Assume that there exists an $\epsilon' \in \R_+$ for which $| \lambda - 
\lambda'| \geq \epsilon'$ holds for all $\lambda' \in \sigma(B)$. Then,
\begin{eqnarray*}
\| B x_0 - \lambda x_0 \|^2 & = & \sum_{i=1}^h |\lambda -\lambda_i |^2 \,
| \langle x_i , x_0 \rangle |^2 \\
& \geq & (\epsilon')^2 \sum_{i=1}^h | \langle x_i , x_0 \rangle |^2 \\
& = & (\epsilon')^2 \, \| x_0 \|^2 \\
& > & (\epsilon')^2 (1 - \epsilon)^2 
\end{eqnarray*}
and we obtain the contradiction $\| B x_0 - \lambda x_0 \| > \epsilon' (1 - \epsilon) >
1/m$. There thus exists a $\lambda' \in \sigma(B)$ for which $\lambda \approx 
\lambda'$ holds. $\diamondsuit$ 
\\*[.3cm]
{\bf Proposition 2:} 
\begin{enumerate}
\item Let $B$ be a nonstandard extension of a self-adjoint operator $A$ as constructed
in Sec. \ref{SOEX}, then
for each $\lambda \in \sigma(A)$ there exists a $\lambda' \in \sigma(B)$ for which
$\lambda \approx \lambda'$ holds, i.e. $\sigma(A) \subset \, ^o\NS( \sigma(B) )$.
\item Let $B$ be a finitely-bounded normal hyperfinite-rank operator, 
then $\sigma(^o B) = \, ^o \sigma(B)$.
\end{enumerate}
{\em Proof}: \\
1. Let $\lambda \in \sigma(A)$, then for each $n \in \N$ there exists
an $x_n \in {\cal M}$ for which $\|x_n\| =1$ and $\| (\lambda - A) x_n \| <
1/n$ holds. Since ${\cal M}$ is externally contained in $H$, we obtain 
$\| (\lambda  - B) \, ^\star x_n \| \leq 1/n$ for each $n \in \N$, and by lemma 1 there
exists a $\lambda' \in \sigma (B)$ with $\lambda \approx \lambda'$. \\
2. We note that this statement can be proved also with the help of lemma 1.
However, the statement is proved in a more general form in \cite{wolff_nsa97},
and we omit therefore the proof. $\diamondsuit$ \\[.3cm] 
The first part of proposition 2 shows that the spectrum of $A$
is approximated well by the eigenvalues of its nonstandard extension $B$. 
We can draw from this result a remarkable conclusion.
Let $\lambda \in \sigma(A)$, then there exists an eigenvalue $\lambda' \approx \lambda$ 
of $B$. Since $B$ has a complete set of (normed) eigenvectors, there exists an $x \in H$
for which $B x = \lambda' x$ holds. For each $y \in {\cal M}$ we obtain thus
\begin{equation}
\langle ^o x , A y \rangle = \, ^o \langle x, B \, ^\star y \rangle = \st(
\lambda') \, ^o\langle x, \, ^\star y \rangle = \lambda \langle ^o x, y \rangle \, .
\end{equation}
Equations of this type can usually be formulated only as eigenfunctional equations
in super-Hilbert-space formalisms. The treatment of continuum states is therefore more
complicated than the treatment of bound states in these frameworks. In a hyperfinite-dimensional
space we can however treat both types of states on an equal footing. We note that
work in this direction is also presented in \cite{far75:177} where the concept of ultra 
eigenvectors is introduced. The concept leads to similar equations, but ultra eigenvectors are not 
necessarily eigenvectors of a nonstandard extension. In particular, we generally do not obtain
an eigenvector basis, which is simpler to use in applications as compared to 
projection-valued measures.

Moreover, if the operator $A$ is bounded in proposition 2 then the nonstandard hull $^oB$ of
$B$ is a bounded self-adjoint operator on $^oH$. This case is extensively studied in  
\cite{ber66:421, ber72:419}. In particular, the operator $A$ is then the restriction of $^oB$ to
${\cal H}$, and the projection-valued measure associated with $A$ can be retrieved
with the help of the eigenvectors of $B$. Unfortunately, most of the self-adjoint operators
occurring in applications are unbounded. Also, we do not know how the function calculus of $A$ is
related to the function calculus of $B$, especially when we consider non-continuous functions.
We therefore use a more general approach that is related to nonstandard integration theory.

\section{Loeb-function calculus}
\label{FC}

In our approach to nonstandard function calculus we introduce first projection-valued Loeb 
measures that are closely related to Loeb measures. The relationship is analogous to the
relationship of standard projection-valued measures to finite Borel measures. We use
projection-valued Loeb measures to prove a nonstandard spectral theorem and to establish
the Loeb-function calculus. Finally, we use the results to introduce generalized nonstandard
hulls, which we use in Sec. \ref{QM} for the further discussion.

\subsection{Projection-valued Loeb measures}
\label{PVLM}

Let ${\cal B}$ denote the set of Borel subsets of $\R$, and let 
$P$ be a finite probability measure on ${\cal B}$. The associated probability space
$(\R, {\cal B}, P )$ transfers to a finitely-additive 
internal probability space, $(^\star \R, \, ^\star {\cal B}, \, ^\star P )$.
To be more general, we replace $^\star P$ by a finitely-additive internal 
probability function $\mu$, and consider the probability space 
$(^\star \R, \, ^\star {\cal B}, \mu )$.
An important result of nonstandard measure theory is the construction of a Loeb
(probability) space out of $(^\star \R, \, ^\star {\cal B}, \mu )$ \cite{ros_nsa97}, i.e.
there exists a standard ($\sigma$-additive) probability space
$(^\star \R, \, ^\star{\cal B}_L, \, \mu_L )$ such that:
\begin{enumerate}
\item $^\star{\cal B}_L$ is a $\sigma$-algebra with
$^\star{\cal B} \subset \, ^\star{\cal B}_L \subset {\cal P(^\star \R)}$.
\item $^\circ \mu = \mu_L$ on $^\star{\cal B}$.
\end{enumerate}
The sets $\Omega \in \, ^\star{\cal B}_L$ are called Loeb-measurable, and $\mu_L$ is
called a Loeb measure. The $\sigma$-algebra $^\star{\cal B}_L$ is however related to
the finitely-additive internal probability function $\mu$, i.e. we should rather
write $^\star{\cal B}_L(\mu)$. To obtain a more general setting we introduce
a smaller $\sigma$-algebra. Let $\Lambda$ denote the set of 
finitely-additive internal probability functions on $^\star{\cal B}$. The
intersection ${\cal A} = \bigcap_{\mu \in \Lambda} \, ^\star{\cal B}_L(\mu)$ 
is a $\sigma$-algebra, and contains
the universally Loeb-measurable sets. For each $\mu \in \Lambda$
the space $(^\star \R, \, {\cal A}, \, \mu_L )$ is a standard
probability space. Moreover, $^\star{\cal B} \subset {\cal A}$ and 
$^\circ \mu = \mu_L$ on $^\star{\cal B}$.

Let $\chi_\Omega$ denote the characteristic function of a set $\Omega \in \, ^\star{\cal B}$.
We may, for example, use the eigensystem of a normal hyperfinite-rank operator $B$ to define a 
finitely-additive internal projection-valued probability function: 
\begin{equation}
E_\Omega = \chi_\Omega(B) = \sum_{\lambda_i \in \Omega} |x_i \rangle \langle x_i|
= \sum_{i=1}^h \chi_\Omega(\lambda_i) \, |x_i \rangle \langle x_i|
\quad (\Omega \in \, ^\star{\cal B}).
\end{equation}
We note that $\{E_\Omega\}$ is the projection-valued $^\star$measure of $B$.
Generally, if we assume a finitely-additive internal projection-valued probability function 
$\{E_\Omega \}_{\Omega \in \, ^\star{\cal B}}$, e.g. a projection-valued $^\star$measure
that is associated with an internal normal operator, 
then we can define for each normed vector $x \in H$, $\|x\|=1$, a probability function 
$\mu^{(x)}$ on $^\star{\cal B}$:
\begin{equation}
\mu^{(x)}(\Omega) = \|E_\Omega x \|^2 = \langle x, E_\Omega x \rangle \quad
(\Omega \in \, ^\star{\cal B} ).
\end{equation}
We denote the associated Loeb measure by $\mu^{(x)}_L$. Moreover, we introduce 
complex-valued Loeb measures. For $x,y \in \fin(H)$ let
\begin{equation}
\mu^{(x,y)}(\Omega) = \langle x, E_\Omega y \rangle 
\quad (\Omega \in \, ^\star{\cal B} ).
\end{equation}
Since $E_\Omega $ is a projection, $\mu^{(x,y)}$ can be decomposed into
four finitely-additive positive internal functions on $^\star{\cal B}$ by 
polarization \cite{ree72}:
\[
\mu^{(x,y)} = \sum_{k=1}^4 a_k \nu^{(k)} , \quad a_k \in \, ^\star \C .
\]
If $\nu^{(k)} \neq 0$ we can use the normalisation $\nu^{(k)}(^\star \R) =1$
without any restriction. Each non-zero $\nu^{(k)}$ can be extended to an ordinary
Loeb measure $\nu^{(k)}_L$. If $\mu^{(x,y)} \neq 0$ we can thus construct a 
finite complex-valued Loeb measure, $\mu_L^{(x,y)}$, that is the sum of up to
four ordinary Loeb measures multiplied by appropriate complex factors. For the
sake of completeness we define further $\mu_L^{(x,y)} = 0$ if $\mu^{(x,y)}=0$.

The family of nonstandard hulls $\{^oE_\Omega \}_{\Omega \in \, ^\star{\cal B}}$ 
is a family of projections on $^oH$, and $\mu_L^{(x,y)}(\Omega) = \langle ^ox, \, ^oE_\Omega
\, ^oy \rangle$ for all $x, y \in \fin(H)$, $\Omega \in \, ^\star{\cal B}$. This result
motivates us to extend the definition of $^oE_\Omega$ to all sets $\Omega \in {\cal A}$.
For $\Omega \in \, ^\star{\cal B}$ we denote the range of 
$^oE_\Omega$ by $R(\Omega)$, which is a closed subspace. For $\Omega \in {\cal A}$
let $^\star{\cal B}_\Omega = \{ \Omega' \in \, ^\star{\cal B}: 
\, \Omega' \subset \Omega \}$. We extend $R(\Omega)$ to ${\cal A}$ by
\begin{equation}
R(\Omega) = \rm{cl} \left( \bigcup_{\Omega' \in \, ^\star{\cal B}_\Omega} R(\Omega') \right),
\end{equation}
and define $^oE_\Omega$ as the projection of $^oH$ onto $R(\Omega)$.
Using a statement in \cite{kad83} on families of projections
we conclude for $x \in H$, $\|x\|=1$, 
\begin{equation}
\langle ^o x, \, ^oE_\Omega \, ^o x \rangle = 
\sup_{\Omega' \in \, ^\star{\cal B}_\Omega} \langle ^o x, \, ^oE_{\Omega'} 
\, ^o x \rangle = \sup_{\Omega' \in \, ^\star{\cal B}_\Omega}
\mu_L^{(x)} (\Omega') = \mu_L^{(x)} (\Omega). 
\end{equation}
\\*[.3cm]
{\bf Theorem 2:} Let $\{E_\Omega \}_{\Omega \in \, ^\star{\cal B}}$
be a finitely-additive internal projection-valued probability function, then
$\{^oE_\Omega\}_{\Omega \in {\cal A}}$ defines a projection-valued Loeb measure. \\[.15cm]
{\em Proof}: We show that the family $\{^oE_\Omega\}_{\Omega \in {\cal A}}$ has the
properties of a projection-valued measure:\\
(a) Assume $\Omega_1, \Omega_2 \in {\cal A}$, and $\Omega_1 \cap \Omega_2 = 
\emptyset$, then $\Omega_1' \cap \Omega_2' = \emptyset$ and
$R(\Omega_1') \bot  R(\Omega_2')$ for all
$\Omega_1' \in \, ^\star{\cal B}_{\Omega_1}$, $\Omega_2' \in \, ^\star{\cal B}_{\Omega_2}$. Thus, 
$R(\Omega_1') \bot  R(\Omega_2)$ for all
$\Omega_1' \in \, ^\star{\cal B}_{\Omega_1}$, $R(\Omega_1) \bot  R(\Omega_2)$, and
$^oE_{\Omega_1} \, ^oE_{\Omega_2} = \, ^oE_{\Omega_2} \, ^oE_{\Omega_1} = 0$. \\
(b) Assume $\Omega_1, \Omega_2 \in {\cal A}$, and $\Omega_1 \subset \Omega_2$, then
$R(\Omega_1) \subset R(\Omega_2)$, and 
$^oE_{\Omega_1} \, ^oE_{\Omega_2} = \, ^oE_{\Omega_2} \, ^oE_{\Omega_1} = \, ^oE_{\Omega_1}$
\cite{kad83}. \\
(c) Assume $\Omega_n \in {\cal A}$, $n \in \N$, and $\Omega_n \cap \Omega_m = \emptyset$ 
if $n \neq m$. Let $\Omega = \bigcup_n \Omega_n$, and let $P_N = \sum_{n=1}^N \, 
^oE_{\Omega_n}$. $P_N$ is a projection since $P_N$ is self-adjoint and $P_N^2 = P_N$
by (a) \cite{ree72}. For $x \in H$, $\|x\|=1$, we obtain by (b)
\begin{eqnarray*}
\| (^oE_\Omega - P_N) \, ^o x \|^2 & = & \langle ^o x, (^oE_\Omega -P_N) \, ^o x \rangle \\
& = & \mu_L^{(x)} (\Omega) - \sum_{n=1}^N \mu_L^{(x)} (\Omega_n) , \\
\lim_{N \to \infty} \| (^oE_\Omega - P_N) \, ^o x \|^2 & = & 0 .
\end{eqnarray*}
(d) Assume $\Omega_1, \Omega_2 \in {\cal A}$. Using (c) we obtain $^oE_{\Omega_1 \cap \Omega_2}
+ \, ^oE_{\Omega_1 \backslash \Omega_2} = \, ^oE_{\Omega_1}$, and using (a), (b) we obtain
$^oE_{\Omega_1} \, ^oE_{\Omega_2} = (^oE_{\Omega_1 \cap \Omega_2}
+ \, ^oE_{\Omega_1 \backslash \Omega_2}) \, ^oE_{\Omega_2} = \, ^oE_{\Omega_1 \cap \Omega_2} =
\, ^oE_{\Omega_2} (^oE_{\Omega_1 \cap \Omega_2}
+ \, ^oE_{\Omega_1 \backslash \Omega_2}) = \, ^oE_{\Omega_2} \, ^oE_{\Omega_1}$.
$\diamondsuit$

\subsection{Spectral theorem}
\label{SPTH}

We have seen in Sec. \ref{PVLM} that a projection-valued $^\star$measure extends to a 
projection-valued Loeb measure $\{^oE_\Omega\}_{\Omega \in {\cal A}}$ on $^oH$. We now use
this result to formulate a nonstandard spectral theorem. We note that the theorem is obtained
not by transfer of a standard spectral theorem. It is rather an analogue associated with Loeb 
integration theory.

We call a complex-valued function $f: \, ^\star\R \to \C$ ${\cal A}$-measurable or Loeb measurable
if $f^{-1} (\Omega) \in {\cal A}$ for each Borel set $\Omega \subset \C$. 
We note that $\st(\cdot)$ is ${\cal A}$-measurable \cite{ros_nsa97}. 
\\[.3cm]
{\bf Theorem 3:} Let $B$ be an internal self-adjoint operator 
on an internal Hilbert space $H$, let $\{^oE_\Omega\}_{\Omega \in {\cal A}}$ be the
projection-valued Loeb measure associated with the projection-valued $^\star$measure of $B$,
and let $f$ be a complex-valued ${\cal A}$-measurable function, then 
$f(B) = \int f \, d\, ^oE$ is a normal operator on $^oH$, and 
$\sigma(f(B)) \subset \rm{cl} (f(\sigma(B)))$.
\\*[.15cm]
{\em Proof}: First we note that if two vectors $x,y \in \fin(H)$ are approximately equal, 
$x \approx y$, then the associated Loeb measures are equal, $\mu_L^{(x)} = \mu_L^{(y)}$.
Consequently, for $x,y \in \, ^oH$ the Loeb measure $\langle x, \, ^oE_\Omega y \rangle$
($\Omega \in {\cal A}$) is well defined, and we use the notation 
$\mu_L^{(x,y)} = \langle x, \, ^oE_\Omega y \rangle$ in this proof.
The domain of $f$ is given by
\begin{equation}
D(f(B)) = \{ x \in \, ^oH: \, \int |f|^2 \, d\mu_L^{(x)} < \infty \}.
\end{equation}
For $n \in \N$ let $\Omega_n = \{ r \in \, ^\star\R: \, |f(r)| \leq n \}$. Each $\Omega_n$ is
measurable, $\Omega_n \in {\cal A}$. Let $x \in \, ^oH$, and let $x_n = \, ^oE_{\Omega_n} x$,
then $x_n \in D(f(B))$ for all $n \in \N$, and $x = \lim_n x_n$. Hence,
$\rm{cl}(D(f(B)) = \, ^oH$. Let $f(B)^\dagger$ denote the adjoint of $f(B)$, and let
$z^\ast$ denote the complex conjugate of $z \in \C$. For
$x \in \, D(f(B))$, $y \in D(f(B)^\dagger)$ we obtain
\begin{eqnarray}
\langle x, f(B)^\dagger y \rangle & = & \langle f(B) x, y \rangle 
 = \langle y, f(B) x \rangle^\ast 
 = \left( \int f \,d \mu_L^{(y,x)} \right)^\ast \\
\nonumber
& = & \int f^\ast \,d \mu_L^{(x,y)} 
 = \langle x, f^\ast(B) y \rangle,
\end{eqnarray}
i.e. $f(B)^\dagger = f^\ast(B)$, $D(f(B)^\dagger) = D(f(B))$, and $f(B)$ is normal.\\
Let $\lambda \in \sigma(f(B))$, and let
$\epsilon > 0$. Since $f(B)$ is normal there exists an $x \in \, ^oH$ for which 
$\|x\|=1$ and 
\[
\| (f(B) - \lambda) x\|^2 = \int |f - \lambda|^2 \,  d\mu^{(x)}_L < \epsilon^2
\]
holds. Let $\Omega_\epsilon = \{r \in \, ^\star\R: \, |f(r) - \lambda| < \epsilon \}
= f^{-1}( \{r \in \R: \, |r - \lambda| < \epsilon \})$, 
then $\mu^{(x)}_L (\Omega_\epsilon) > 0$, and thus $^oE_{\Omega_\epsilon} \neq 0$.
Moreover, there exists an $\Omega'_\epsilon \in \, ^\star{\cal B}_{\Omega_\epsilon}$
for which $\mu^{(x)}_L (\Omega'_\epsilon) > 0$ holds. Thus, $E_{\Omega'_\epsilon} \neq 0$
and $\Omega'_\epsilon \cap \sigma(B) \neq \emptyset$. Choose $\omega_\epsilon \in
\Omega'_\epsilon \cap \sigma(B)$, then $|f(\omega_\epsilon ) - \lambda| < \epsilon$, and
thus $\lambda \in \rm{cl} (f(\sigma(B)))$. $\diamondsuit$ \\*[.3cm]
{\bf Corollary 1:} Assume the conditions of theorem 3. If 
$B$ is additionally a hyperfinite-rank operator, then 
\begin{enumerate} 
\item
$\sigma(f(B)) = \rm{cl} (f(\sigma(B)))$,
\item
$\sigma(^o g(B)) = \, ^o g(\sigma(B))$ for 
finitely bounded $g \in \, ^\star{\cal B}(\R)$.
\end{enumerate}
{\em Proof}: \\
1. Let $\lambda \in \sigma(B)$, and let $x$ be a corresponding eigenvector, then
$^o E_{\{ \lambda \}} \, ^ox = \, ^ox$, $f(B) \, ^ox = f(\lambda) \, ^ox$, and
$f(\lambda) \in \sigma(f(B))$. 
Since $\sigma(f(B))$ is closed we obtain $\rm{cl} (f(\sigma(B))) \subset \sigma(f(B))$. 
The assertion follows now from theorem 3. \\
2. Let $g \in \, ^\star{\cal B}(\R)$ be finitely bounded. Since $B$ is a hyperfinite-rank
operator we obtain $\sigma(g(B)) = g (\sigma(B))$. Thus, $\sigma(^og(B)) = \,
^o\sigma(g(B)) = \, ^og (\sigma(B))$ by proposition 2. $\diamondsuit$ \\[.15cm]
If we consider in theorem 3 a real-valued function $f$, then we obtain a self-adjoint
operator $f(B)$. For this operator, the standard spectral theorem yields the standard spectral
representation. Since theorem 3 states an alternative nonstandard representation of $f(B)$ 
we clarify now the relationship between both representations.
\\[.3cm]
{\bf Theorem 4:} Let $B$ be an internal self-adjoint operator 
on an internal Hilbert space $H$,
let $f$ be a real-valued ${\cal A}$-measurable function, and let $g$ be a Borel
function, then $g(f(B)) = (g \circ f) (B)$.
\\*[.15cm]
{\em Proof}: For a Borel set $\Omega \in {\cal B}$ let $F_\Omega = \, ^oE_{f^{-1}(\Omega)}$.
$F_\Omega $ defines a projection-valued measure, and for $x \in \fin (H)$ we obtain
\[
\mu_{f,L}^{(x)} (\Omega) = \mu_L^{(x)} (f^{-1}(\Omega)) = 
\langle ^o x, \, ^oE_{f^{-1}(\Omega)} \, ^o x \rangle =
\langle ^o x, \, F_\Omega \, ^o x \rangle
\]
Let us assume first that $g$ is bounded, then we obtain for $\Omega \in {\cal B}$ 
\[
\int\limits_\Omega g \, d \mu_{f,L}^{(x)} = \int\limits_{f^{-1}(\Omega)} g \circ f 
\, d\mu_L^{(x)} \,.
\]
We conclude that $\int g \, dF = \int g \circ f \, d\, ^oE$.
If we apply pointwise-convergence arguments the latter equation is valid for any Borel function $g$.
In particular, for $g = \mbox{id}$ we obtain $f(B) = \int \omega \, dF_\omega$, which
is just the standard spectral representation. Since $\int g \, dF = g(\int \omega \, 
dF_\omega) = g(f(B))$ we obtain the statement of the theorem. $\diamondsuit$ \\[.15cm]
We note that theorem 4 reveals the relationship between the standard and the nonstandard spectral 
representation of the self-adjoint operator $f(B)$: For each Borel set $\Omega$ we obtain
$F_\Omega := \chi_\Omega(f(B)) = \int \chi_\Omega \circ f \, d\, ^oE = 
\, ^oE_{f^{-1}(\Omega)}$, and $f(B) = \int \omega \, dF_\omega$.

\subsection{Nonstandard hulls}
\label{NH}

As pointed out in Sec. \ref{BF} operational nonstandard hulls are well-known for finitely-bounded
internal operators. For infinite internal operators, however, ambiguities occur if we consider
standard parts: Let $B$ be a hyperfinite-rank operator that has an infinite eigenvalue $\lambda$,
let $x$ be the corresponding normed eigenvector, and let $y = x/\lambda$, then
$^oy = 0$ but $^o(By) = \, ^ox \neq 0$. We see by this example that for infinite internal
operators an equation like $^oB \, ^ox = \, ^o(Bx)$ can generally not be true.
The definition of nonstandard hulls for infinite internal
operators is thus not straightforward. Nevertheless, if we consider self-adjoint 
internal operators then we can use the Loeb-function calculus for a sensible
definition. The principal idea is to project those vectors out that cause the ambiguities.
To arrive at a sensible definition we consider first finitely-bounded internal operators.
Moreover, let ${\cal B}(\R)$ be the set of complex-valued Borel functions on $\R$.
\\*[.3cm]
{\bf Proposition 3:} 
Let $B$ be an internal self-adjoint operator 
on an internal Hilbert space $H$, let $\{^oE_\Omega\}_{\Omega \in {\cal A}}$ be the
projection-valued Loeb measure associated with the projection-valued $^\star$measure of $B$,
and let $f \in \, ^\star{\cal B}(\R)$ be finitely bounded, then
\begin{equation}
\langle ^ox, \, ^of(B) \, ^oy \rangle =  \st \left( \int\limits f(\omega) \, 
d \langle x, E_\omega y \rangle \right)
= \int \, ^of \, d\mu_L^{(x,y)} \quad (x,y \in \fin(H)).
\end{equation}
{\em Proof}: We assume first that $f$ is real-valued. 
The generalization to complex-valued functions $f$ is
straightforward. \\
For $n \in \, ^\star\N$, $k \in  \, ^\star\Z$ let 
$S_{k,n} = \{ r \in  \, ^\star\R: \, k/n \leq f(r) < (k+1)/n \}$, and let
$I_n = \{k \in  \, ^\star\Z: \, S_{k,n} \neq \emptyset \}$. Since $f$ is finitely bounded
$I_n$ is hyperfinite. Moreover, $^\star\R = \bigcup_{k \in I_n} S_{k,n}$. Define
$f_n = \sum_{k \in I_n} \frac{k}{n} \, \chi_{S_{k,n}}$, then $f_n \uparrow f$, and
$|f_n(r) - f(r)| < 1/n$ for all $r \in  \, ^\star\R$. Moreover, 
$^of_n \uparrow \, ^of$ ($n \in \N$), and
\begin{eqnarray*}
\int \, ^of \, d\mu_L^{(x,y)} & = & \lim_{n \in \N} \int \, ^of_n \, d\mu_L^{(x,y)} 
\quad (x,y \in \fin(H)) \\
& = & \lim_{n \in \N} \sum_{k \in I_n} \frac{k}{n} \, \mu_L^{(x,y)}(S_{k,n}) \\
& = & \lim_{n \in \N} \, ^o\sum_{k \in I_n} 
\int \frac{k}{n} \, \chi_{S_{k,n}}(\omega)  \, d \langle x, E_\omega y \rangle \\
& = & \int\limits^{o \quad} \, f_n(\omega) \, d \langle x, E_\omega y \rangle 
 \quad (n \in \, ^\star\N \backslash \N) \\
& = & \int\limits^{o \quad} \, f(\omega) \, d \langle x, E_\omega y \rangle \\
& = & ^o \langle x, \, f(B) y\rangle = \langle ^o x, \, ^o f(B) \, ^o y\rangle .
\end{eqnarray*}
We note that we use $f_n (r) \approx f(r)$ for $n \in \, ^\star\N \backslash \N$
in the calculation. $\diamondsuit$ \\[.3cm]
We note that if the operator $B$ is finitely bounded in proposition 3 then
the spectrum of $B$ consists only of near-standard points, $\sigma(B) \subset
\fin(^\star \R)$, and thus $^oB = \int_{\fin(^\star \R)} \st \, d^oE$. Moreover,
the last equation provides a correct approach for an extended definition
of nonstandard hulls.
\\*[.3cm]
{\bf Definition 1:} Let $B$ be an internal self-adjoint operator 
on an internal Hilbert space $H$, and let $\{^oE_\Omega\}_{\Omega \in {\cal A}}$ be the
projection-valued Loeb measure associated with the projection-valued $^\star$measure of $B$.
The nonstandard hull of $B$ is given by
\[
^oB = \int_{\fin(^\star \R)} \st \, d^oE. 
\]
This definition avoids the ambiguities mentioned above since the critical vectors, i.e.
the vectors belonging to the range of $^oE_{\fin(^\star \R) \setminus ^\star \R}$, are
projected out. In particular, if $x$ is an eigenvector of $B$ in definition 1, then
$^oB \ ^ox = 0$ if $x$ belongs to an infinite eigenvalue. We note that
the statement $(\forall x \in \fin(H) ) \, ^oB \, ^ox = \, ^o(Bx)$ holds for finitely bounded 
operators but not for infinite operators in definition 1. Generally, we obtain 
for $x, Bx \in \fin(H)$ rather 
\begin{equation}
^oB \, ^ox = \, ^oE_{\fin(^\star \R)} \, ^o(Bx).
\end{equation}

\section{Nonstandard quantum mechanics}
\label{QM} 

\subsection{Operational extensions}
\label{OE}

We continue now our discussion started in Sec. \ref{BF}. In particular, we consider a
dense subset ${\cal M}$ of a Hilbert space ${\cal H}$, which is externally contained in a
hyperfinite dimensional space $H$, and for which ${\cal M} \subset H \subset 
\, ^\star{\cal M} \subset \, ^\star{\cal H}$ holds. We note that ${\cal H} \subset
\, ^oH$ by proposition 1. We have seen in Sec. \ref{BF} 
that each bounded linear operator on ${\cal H}$ can be extended by a 
finitely-bounded hyperfinite-rank operator on $H$, 
and that the extension is based on nonstandard hulls. In Sec. \ref{FC} we have worked 
out the connection of nonstandard hulls to the Loeb-function calculus. In particular,
we have introduced nonstandard hulls for infinite self-adjoint internal operators.
Now we use these results to generally establish the extension of 
standard self-adjoint operators to self-adjoint hyperfinite-rank operators.

Let $A$ be a self-adjoint operator on ${\cal H}$, then $A' = \tan^{-1} (A)$ is a 
bounded self-adjoint operator on $H$. The hyperfinite-rank
operator $B' = O \,^\star A' O$ is a nonstandard extension of $A'$, i.e.
$^oB'|_{\cal H} = A'$. We note that $O$ denotes the projection of ${\cal \,
^\star H}$ onto $H$. Moreover,
we note that $\chi_{(-\pi/2 , \pi/2)}(A') = \chi_{\R} (A) = 1$ and that 
$A = \tan(A')$. For $r \in \R$ let $\tg(r) = \tan( r \cdot \chi_{(-\pi/2 , \pi/2)}(r))$.
Since $^oB'|_{\cal H} = A'$ we obtain $\tg(A') = \tg(^oB')|_{{\cal D}(\tg(A'))}$.
Since $\st \circ \tg = \tg \circ \, \st$ on $\fin(^\star \R)$ we get further
$\tg(^oB') = \, ^o\tg (B')$ by proposition 3 and theorem 4, and thus
$A = \tg(A') = \, ^o(\tg(B'))|_{{\cal D}(A)}$. 

Our construction shows that for any standard self-adjoint operator $A$ there exists a self-adjoint
hyperfinite-rank operator $B$ whose nonstandard hull extends $A$, i.e.
$A =  \, ^oB|_{{\cal D}(A)}$. In particular, we obtain $\sigma(A) \subset \sigma(^oB) = 
\mbox{cl} ( ^o\NS(\sigma(B))$ by corollary 1. For each 
$\lambda \in \sigma(A)$ there exists thus a $\lambda' \in \sigma(B)$ for which
$\lambda \approx \lambda'$ holds (c.f. Sec. \ref{SPEX}). 
\\*[.3cm]
{\bf Definition 2:} Let $A$ be a self-adjoint operator on a Hilbert space ${\cal H}$. Let $H$
be an internal Hilbert space for which ${\cal H} \subset \, ^oH$ holds and let
$B$ be an internal self-adjoint operator on $H$. $B$ is called a nonstandard extension of $A$ if 
$A =  \, ^oB|_{{\cal D}(A)}$.\\[.15cm]
We note that the nonstandard extension of $A$ is not unique. We discuss now how the 
function calculus of $A$ is related to the function calculus of its 
nonstandard extensions. \\[.3cm]
{\bf Proposition 4:} Let $A$ be a standard self-adjoint operator and let $B$ 
be a nonstandard extension of $A$. 
Then, for each real-valued Borel function $g$ there exists an internal function $f$ for which
$g(A) x = \, ^o( f(B)  \, ^\star x ) = \, ^of(B)x$ holds for all $x \in {\cal D}(g(A))$. 
In particular, $f(B)$ is a nonstandard extension of $g(A)$. \\[.15cm]
{\em Proof}: Let $g$ be a real-valued Borel function and 
let $\{^oE_\Omega\}_{\Omega \in {\cal A}}$ be the
projection-valued Loeb measure associated with the projection-valued $^\star$measure of $B$.
Since $g(A) = g(^oB)|_{{\cal D}(g(A))}$ we conclude from theorem 4 that
$g(A) x =  \int_{\fin(^\star \R)} (g \circ \, \st) d\, ^oE x$ ($x \in {\cal D}(g(A))$). 
Moreover, for $x \in {\cal D}(g(A))$ let $\mu_L^{(x)}$ be the Loeb measure for which 
$\langle x, \, ^oE_\Omega x \rangle = \mu_L^{(x)}(\Omega)$ $(\Omega \in {\cal A})$ holds. 
For each Loeb measure $\mu_L^{(x)}$ there exists an internal function $f_x$ for which 
$^of_x = g \circ \, \st$ holds $\mu_L^{(x)}$ almost everywhere \cite{ros_nsa97}, i.e. 
$\int_{\fin(^\star \R)} ( ^of_x - g \circ \, \st)^2 \, d\|^oEx\|^2 = 0$. 
For $n \in \N$ let 
\[
f_{x,n} (\lambda) := 
\left\{ \begin{array}{c@{\quad,\quad}c} f_x(\lambda) &  |f_x(\lambda) | \leq n  \\ 
0 & \mbox{else.} \end{array} \right. ,
\]
then each $f_{x,n}$ is a finite internal function and $g(A) x = g(^oB) x =$ \\
$\lim_n \, ^of_{x,n}(B) x = \lim_n \, ^o(f_{x,n}(B) \, ^\star x)$. Hence, the internal sets 
\[
\Omega_{x,n} = \{ f \in \, ^\star (\C^{\R}) \, : \, \| ^\star(g(A) x) - f(B) \, ^\star x
\| < 1/n \} 
\]
are non-empty for each $x \in {\cal D}(g(A))$, $n \in \N$. We note that $\Omega_{x,n} \subset
\Omega_{x,m}$ for $n > m$. Moreover, for $x_1, ..., x_m \in {\cal D}(g(A))$ we consider the
Loeb measure $\nu_L = \mu_L^{(x_1)} + ... + \mu_L^{(x_m)}$. For $\nu_L$ there exists also
an internal function $f$ for which $^of = g \circ \, \st$ holds $\nu_L$ almost 
everywhere \cite{ros_nsa97}. In particular, $^of = g \circ \, \st$ holds 
$\mu_L^{(x_k)}$ almost everywhere for each $1 \leq k \leq m$. 
If we define $f_n$ with the help of $f$ analogously as we have defined $f_{x,n}$ 
with the help of $f_x$ then $g(A) x = g(^oB) x_k = \lim_n \, ^o
f_n(B)x_k = \lim_n \, ^o(f_n(B) \, ^\star x_k)$ for each $1 \leq k \leq m$, and thus
$\Omega_{x_1,n} \cap ... \cap \Omega_{x_m,n} \neq \emptyset$. The collection of internal sets
$(\Omega_{x,n})_{x \in {\cal D}(g(A)), n \in \N}$ has thus the finite-intersection
property, and by polysaturation there exists an internal $f \in 
\bigcap_{x \in {\cal D}(g(A)), n \in \N} \Omega_{x,n}$. In particular, 
we obtain $g(A) x = \, ^o( f(B)  \, ^\star x )$ for all $x \in {\cal D}(g(A))$. \\
Assume $x \in {\cal D}(g^2(A))$ and let $y = g(A)x$, then $^\star y \approx f(B) \, ^\star x$
and $^\star(g^2(A) x) = \, ^\star(g(A) y) \approx f(B) \, ^\star y$. Since $f(B) \, ^\star y$
is finite we obtain $y = Py$ for the projection $P = \, ^o\chi_{\fin(^\star\R)}(f(B))$. Thus,
$y = \, ^o( f(B) \, ^\star x) = P \, ^o( f(B) \, ^\star x) = \, ^of(B) x$. Since $g(A)$ is
essentially self-adjoint on ${\cal D}(g^2(A))$ we obtain $g(A) = \, ^of(B)|_{{\cal D}(g(A))}$.
$\diamondsuit$ 
\\[.15cm]
We note that the internal function in proposition 4 is not uniquely determined. 
In fact, if $B$ is of hyperfinite rank we can even choose a $^\star$polynomial $p$ for which 
$p(\lambda) = f(\lambda)$ holds for all eigenvalues $\lambda$ of $B$, i.e. $f(B) = p(B)$. 
We can also restrict $f$ to be a $^\star$continuous function or a $^\star$Borel function in 
proposition 4. We note also that if in proposition 4 the function $g$ is bounded and continuous
($\mu_L^{(x)}$ almost everywhere for all $x \in {\cal H}$) then we may simply choose 
$f = \, ^\star g$. 
\\[.3cm]
{\bf Proposition 5:} Let $A$ be a standard self-adjoint operator and let $B$ 
be a nonstandard extension of $A$. 
Let $(F_\Omega)_{\Omega \in {\cal B}}$ be the projection-valued measure associated with $A$
and let $(E_\Omega)_{\Omega \in {\cal \,^\star B}}$ be the projection-valued $^\star$measure 
associated with $B$, then for each $\Omega \in {\cal B}$ there exists an $\Omega' \in 
{\cal \,^\star B}$ for which $F_\Omega x = \, ^oE_{\Omega'} x$ holds for all $x \in
{\cal H}$. \\*[.15cm]
{\em Proof}: Let $x \in {\cal D}(A)$, then $F_\Omega x = \chi_\Omega(A) x =
\chi_\Omega(^oB) x = \, ^oE_{\st^{-1}(\Omega)} x$ by theorem 4. 
Let $\mu_L^{(x)}$ be the Loeb measure associated with
$\langle x, \, ^oE_\Omega' x \rangle$ $(\Omega' \in {\cal \,^\star B})$. There exists
an $\Omega'_x \in {\cal \,^\star B}$ for which $\mu_L^{(x)} (\st^{-1}(\Omega)
 \Delta \Omega'_x) = 0$ holds \cite{ros_nsa97} ($\Omega \Delta \Omega' = ( \Omega \setminus
 \Omega') \cup ( \Omega' \setminus \Omega)$). We get 
in particular $^oE_{\Omega'_x} x = \, ^oE_{\st^{-1}(\Omega)} x $.  Let 
\[
\Gamma_{x,n} = \{ \Omega' \in {\cal \,^\star B} \, : \, \| ( E_{\Omega'} - E_{\Omega'_x} ) \,
^\star x \| < 1/n \} .
\]
The sets $\Gamma_{x,n}$ are internal and $\Gamma_{x,n} \subset \Gamma_{x,m}$ if $n > m$.
Moreover, for $x_1, ..., x_m \in {\cal D}(A)$ we consider the
Loeb measure $\nu_L = \mu_L^{(x_1)} + ... + \mu_L^{(x_m)}$. For $\nu_L$ there exists also
an $\Omega' \in {\cal \,^\star B}$ for which $\nu_L (\st^{-1}(\Omega) \Delta \Omega') = 0$
holds, i.e. $^oE_{\Omega'} x_k = \, ^oE_{\st^{-1}(\Omega)} x_k $ for $1 \leq k \leq m$.
The collection of internal sets $(\Gamma_{x,n})_{x \in {\cal D}(A), n \in \N}$ 
has thus the finite-intersection property, and by polysaturation there exists an 
$\Omega' \in \bigcap_{x \in {\cal D}(A), n \in \N} \Gamma_{x,n}$. In particular,
$^oE_{\Omega'} x = \, ^oE_{\st^{-1} (\Omega)} x $ for all $x \in {\cal D}(A)$, which proves the
assertion. $\diamondsuit$ 
\\[.15cm]
Let us come back to the discussion at the beginning of this section. We have seen that for 
any self-adjoint operator $A$ on ${\cal H}$ there exist 
self-adjoint hyperfinite-rank operators on $H$ that extend $A$ in the sense of definition 2. 
Proposition 4 and proposition 5 provide further results that we discuss now from a 
physical point of view:
\begin{enumerate}
\item
If we want to model a standard measurement process we 
usually use a probability space $(\R, {\cal B}, \mu)$. The measure $\mu$ is related to
a projection-valued measure of a self-adjoint operator $A$ and a normalized state 
vector $x \in {\cal H}$. Proposition 5 suggests that we could also use a 
$^\star$probability space $(\R, {\cal \, ^\star B}, \mu)$ to model the measurement process.
The $^\star$measure $\mu$ is then related to the projection-valued $^\star$measure of
a self-adjoint hyperfinite-rank extension $B$ of $A$ and the normalized vector $y =
O\, ^\star x / \|O\, ^\star x\|$. We note that $O$ denotes the projection of ${\cal \,
^\star H}$ onto $H$ and that $O\, ^\star x \approx \, ^\star x$.
\item Proposition 4 tells us that the set of $^\star$Borel functions contains all functions
that we need to retrieve standard results. We note that we may restrict ourselves to
any set that contains the $^\star$polynomials, i.e. the $^\star$continuous functions, for
example. In particular, for the time evolution we simply obtain $\exp(-iAt) x = \,
^o( \exp(-iBt) \, y) = \, ^o\exp(-iBt) x$ ($t \in \R$). We note that we use $A$, $B$, $x$, and
$y$ as in 1.
\end{enumerate}
Following these two arguments we may formulate quantum mechanics in a hyperfinite-dimensional
Hilbert space $H$, which extends the standard formulation in a Hilbert space
${\cal H}$. However, our results have a rather general nature so far. To demonstrate how a
concrete formulation can be achieved we discuss Schr\"odinger representations as an example in
Sec. \ref{SO}.

\subsection{Schr\"odinger operators}
\label{SO}

The conventional Schr\"odinger representation of one-particle quantum mechanics
is formulated in ${\cal H} = {\cal L}^2 (\R)$. 
The momentum operator $p$ and the position operator $q$ are the closures
of $i^{-1} d/dx$ and multiplication by $x$ on ${\cal S} (\R)$. ${\cal S} (\R)$ is the space of
functions of rapid decrease on $\R$, which is a domain of essential self-adjointness for $p$ and
$q$. Moreover, Schr\"odinger Hamiltonians are given by $A = p^2 + V(q)$, for which 
$V(\cdot)$ denotes a Borel function. We note that we restrict ourselves to the 
one-dimensional case and that the generalization to multi dimensions is straightforward, 
as we will see. 

Let $\{u_n\}_{n = 0}^\infty$ be the basis of Hermite functions, let $h \in \, ^\star\N \setminus
\N$ be a fixed hypernatural number, and let $H$ be the hyperfinite-dimensional space that has 
the basis $\{^\star u_n\}_{n = 0}^h$. $H$ externally contains the space ${\cal M}$ of finite
linear combinations of $\{u_n\}_{n = 0}^\infty$, which is dense in ${\cal H}$, and
$H \subset \, ^\star{\cal M}$. The triple $({\cal M},{\cal H},H)$ thus realizes the
setting described in Sec. \ref{SOEX}.

We derive now explicit nonstandard extensions of several self-adjoint operators on ${\cal H}$.
For example, let $O = \sum_{n=0}^h | \, ^\star u_n \rangle \langle \, ^\star u_n |$ be 
the projection from $^\star{\cal H}$ onto $H$ associated with the Hermite-function basis 
$\{^\star u_n\}_{n = 0}^h$, then the hyperfinite-rank operator
$P = O\, ^\star p \, O$ has a well-known simple representation, 
and for $x \in {\cal M}$ we obtain $^\star(p x) = P \, ^\star x$. Let 
$(E_\Omega)_{\Omega \in {\cal \,^\star B}}$ be the projection-valued 
$^\star$measure associated with $P$. Since $P^2 \, ^\star x$ is finite we obtain $p x = \, 
^o( P \, ^\star x) =  \, ^oE_{\fin (\R)}\, ^o( P \, ^\star x) = \, ^oP x$ for $x \in {\cal M}$. 
In order to show that $P$ is a nonstandard extension of $p$ we prove that 
$p|_{\cal M}$ is essentially self-adjoint. 

Let $x \in {\cal S}$ then there exists a sequence $(x_n)$ in ${\cal M}$
that converges to $x$ with respect to the topology of ${\cal S}$ \cite{ree72}. 
In particular, $\lim_n \| x - x_n \| = 0$ and $\lim_n \|p(x - x_n)\| = 0$.
Let $\Gamma(p |_{\cal M})$ denote the graph of $p$ restricted to ${\cal M}$.
$(px, x)$ is thus contained in the closure of $\Gamma(p|_{\cal M})$, i.e.
$\Gamma(p|_{\cal S}) \subset \mbox{cl} (\Gamma(p|_{\cal M})) = \mbox{cl} 
(\Gamma(p|_{\cal S}))$. Since $p|_{\cal S}$ is essentially self-adjoint we conclude that 
$p|_{\cal M}$ is also essentially self-adjoint and that
$p = \ ^oP|_{{\cal D}(p)}$. $P$ is thus a 
nonstandard extensions of $p$. The following lemma generalizes our result.\\[.3cm]
{\bf Lemma 2:} Let $H$ be an internal Hilbert space and let ${\cal M}$ be a dense
subset of a standard Hilbert space ${\cal H}$ for which ${\cal M} \subset H \subset 
\, ^\star{\cal M} \subset \, ^\star{\cal H}$ holds. Let $A$ be a self-adjoint 
operator on ${\cal H}$, which is essentially self-adjoint on ${\cal M}$, and let
$O$ be the projection of $^\star{\cal H}$ onto $H$. Then,
$B = O \,\, ^\star A O$ is a nonstandard extension of $A$.
\\[.15cm]
{\em Proof}:
Let $(F_\Omega)_{\Omega \in {\cal \,^\star B}}$ be the projection-valued 
$^\star$measure associated with $B$. Since
\[
\| B \, ^\star x \|^2 = \int\limits_{^\star \R} 
\omega^2 \, d\langle \, ^\star x, F_\omega \, ^\star x \rangle 
\]
is finite for $x \in {\cal M}$ we conclude that
\[
\int\limits_{| \omega | > n} 
|\omega| \, d\langle \, ^\star x, F_\omega \, ^\star x \rangle < \epsilon
\]
for any standard $\epsilon > 0$ and any $n \in \, ^\star \N \setminus \N$. For a fixed
standard $\epsilon > 0$ there exists by the underflow principle an $n \in \N$ for which
\[
\int\limits_{| \omega | > n} 
|\omega| \, d\langle \, ^\star x, F_\omega \, ^\star x \rangle < \epsilon
\]
holds, and thus 
\begin{eqnarray*}
\langle x, \, ^o( B \, ^\star x) \rangle & = & \st \left( \, \int\limits_{^\star \R} 
\omega \, d\langle \, ^\star x, F_\omega \, ^\star x \rangle \right) 
= \lim_{n \to \infty} \, \st \left( \, \int\limits_{| \omega | \leq n} 
\omega \, d\langle \, ^\star x, F_\omega \, ^\star x \rangle \right) \\
& = & \langle x, \, ^oB x \rangle \quad (x \in {\cal M}).
\end{eqnarray*}
Moreover, by polarization we obtain $\langle y, \, ^o(B \, ^\star x) 
\rangle = \langle y, \, ^oB x \rangle$ for $x, y \in {\cal M}$. 
Since $^o( B \, ^\star x) = A x \in {\cal H}$ we get
\[
\| ^o( B \, ^\star x) \| = \sup_{y \in {\cal M}_1} | \langle y, \, ^o(B \, ^\star x) 
\rangle | = \sup_{y \in {\cal M}_1} | \langle y, \, ^oB x \rangle | \leq \| ^oB x \|,
\]
using ${\cal M}_1 = \{ y \in {\cal M} : \, \| y \| =1 \}$. Since 
\[
\| ^o( B \, ^\star x) \| = \| ^oB x \| +  \| ^oF_{^\star \R \setminus \fin(^\star \R)} \, 
^o( B \, ^\star x) \|
\]
we obtain $ ^oF_{^\star \R \setminus \fin(^\star \R)} \, ^o( B \, ^\star x) = 0 $ and
$Ax = \, ^o( B \, ^\star x) = \, ^oB x$ for $x \in {\cal M}$. Since $A|_{\cal M}$ is
essentially self-adjoint we obtain $A = \, ^oB|_{{\cal D}(A)}$. $\diamondsuit$ 
\\[.15cm]
With the help of lemma 2 we gain explicit nonstandard extensions of many self-adjoint
operators on ${\cal H}$. For example, since $q^n|_{\cal S}$ and $p^n|_{\cal S}$ 
($n \in \N$) are
essentially self-adjoint we conclude in the same manner as for $p$ that
$Q^{(n)} = O\, ^\star q^n \, O$ and $P^{(n)} = O\, ^\star p^n \, O$ 
are nonstandard extensions,
respectively. Also, if for a real Borel function $V$ the operator $V(q)|_{\cal M}$  
is essentially self-adjoint, which is for example true if $V$ is bounded, 
then $W = O\, ^\star V(q) \, O$ is a nonstandard extension of $V(q)$. We note that
the matrix elements of $W$ are given by explicit analytic expressions. Moreover,
let us assume that the Schr\"odinger operator $A = p^2 + V(q)$ is essentially 
self-adjoint on ${\cal M}$ and that ${\cal M} \subset {\cal D}(V(q))$. This requirement
is fulfilled, for example, if $V$ is a bounded Borel function (c.f. W\"ust's
theorem \cite{ree75}). Using lemma 2 we conclude that 
$B = O ( ^\star p^2 + \, ^\star V(q)) O = P^{(2)} + W$ is a nonstandard extension of $A$.
Without giving an explicit proof we note that we may even choose
$V \in {\cal L}^2 + {\cal L}^\infty$ as in theorem X.15 of \cite{ree75}. 
Moreover, we note that in this example we simply may add the nonstandard extensions of $p^2$ 
and $V(q)$ to obtain a nonstandard extension of $A$. However, this is generally not valid
since for two internal self-adjoint operators $B_1$ and $B_2$ the sum $^oB_1 + \, ^oB_2$
is not necessarily defined although $^o(B_1 + B_2)$ is always defined.

\section{Application in scattering theory}
\label{ST}

We apply in this section our nonstandard framework to non-relativistic quantum
scattering theory. First, we shortly review the concepts of conventional 
time-dependent scattering theory as it is presented in \cite{ree79}. We show 
then how the concepts fit into the framework, and discuss the physical 
impact. We finally derive explicit expressions for the M\o ller wave operators 
and the S-Matrix.

\subsection{Time-dependent scattering theory}
\label{TDST}

In quantum scattering theory the Hamiltonian of the quantum
system is the sum of a ''free'' Hamiltonian and an interaction
potential, $A = A_0 + V$. If we consider a state $x$ that was prepared
in the remote past then the corresponding free state $x_+$ is given by $x = W^+ x_+
= \lim_{t \to -\infty} e^{iAt}e^{-iA_0t} x_+$. $W^+$ is a M\o ller wave operator. 
Prerequisite is however that $\lim_{t \to -\infty} e^{iAt}e^{-iA_0t} x_+$
exists. Analogously, the free state $x_-$ that looks like $x$ when it is detected
in the far future is given by $x= W^- x_-
= \lim_{t \to \infty} e^{iAt}e^{-iA_0t} x_-$. We note that we use the convention of
time-independent scattering theory 
that $t \to \mp \infty$ refers to $W^\pm$ (c.f. Sec. \ref{TIST}, \cite{ree79}).
The quantum system is complete if $W^+({\cal H}) = W^-({\cal H}) = {\cal H}_{ac}$.
${\cal H}_{ac}$ is the subspace of ${\cal H}$ that is connected to the 
absolutely continuous part of the spectrum of $A$. 

Let us assume now an internal Hilbert space $H$, and self-adjoint
hyperfinite-rank operators $B$ and $B_0$ on $H$. We associate $B_0$ with the free
Hamiltonian of the quantum system. Analogous to the conventional theory we are
interested in the limits $W^\pm x =\lim_{t \to \pm \infty} (^oe^{iBt})(^o e^{-iB_0t}) 
x$ ($x \in \, ^o H$). Since we use NSA we can however 
give the terms ''remote past'' and ''far future''
a quantitative meaning. Let $Y_t = e^{iBt} e^{-iB_0t}$ and let
$L^\pm = \{ x \in \fin(H): \, \lim_{t \to \mp \infty} 
\, ^o(Y_t \, x) \,\, \mbox{exists} \}$. 
We note that the spaces $^oL^\pm$ are closed subspaces
of $^oH$ on which the M\o ller operators $W^\pm$ are defined. We assume in the 
following that the system is ''reasonably'' complete, i.e. that
$W^+(L^+) \cap W^-(L^-)$ is a sensible set of physical states. This assumption is 
justified if we consider the nonstandard extension of a complete standard
quantum system.
\\[.3cm]
{\bf Lemma 3:} $x \in L^\pm$ iff there exist  
infinite hyperreals $T_{\pm,x}$ for which $W^\pm \, ^ox = \, ^o(Y_t\,x)$ holds if $t$
is infinite and $0 < (\mp t) \leq T_{\pm,x}$. \\[.15cm]
{\em Proof}: We carry out only the proof for $L^-$, since the proof for $L^+$ is 
analogous. Let $x \in \fin(H)$ and assume that there exists an infinite
$T_{-,x}$ for which $W^- \, ^ox = \, ^o(Y_t \, x)$ holds if $t$ is infinite and
$0 < t \leq T_{-,x}$. Since $^o(Y_t \, x) \in \, ^oH$ there exists a $y \in \fin(H)$
for which $^oy = \, ^o(Y_t \, x)$ holds. For a (standard) $\epsilon > 0$ let
\[
\Gamma_\epsilon = \{ T \in \, [0,T_{-,x}]: \, 
(\forall t \in [T,T_{-,x}]) \, \| y - Y_tx\| < \epsilon \}.
\]
Since  $T \in \Gamma_\epsilon$ if $T$ is infinite and $T_{-,x} \geq T > 0$ , and since 
$\Gamma_\epsilon$ is internal and non-empty there exists a finite 
$T_\epsilon \in \Gamma_\epsilon$.
Hence, $(\forall t > \, ^oT_\epsilon) \, \|W^- \, ^ox - ^o(Y_t\,x)\| \leq \epsilon$, and
thus $W^- \, ^ox = \, \lim_{t \to \infty} \, ^o(Y_t\,x)$. \\
Now assume $W^- \, ^ox = \lim_{t \to \infty} \,^o(Y_t\,x)$. Since $W^- \, ^ox \in \, ^oH$ 
there exists a $y \in \fin(H)$ for which $^oy = W^- \, ^ox$ holds. 
For each (standard) $\epsilon > 0$ there exists a $T_\epsilon$ for which 
$(\forall t > T_\epsilon) \, \|W^- \, ^ox - \,^o(Y_t\,x)\| < \epsilon$
holds. Let $T'_\epsilon = \max\{ T_\epsilon, 1/\epsilon \}$, and let 
$F_t = \{ s \in \, ^\star \R: \, s \geq t \}$ for $t \in \, ^\star \R$. The set
\[
G_{\epsilon,x} = \{ T \in F_{T'_\epsilon}: \, (\forall t \in [T'_\epsilon, T])
\, \| y - Y_t\,x\| < \epsilon \}
\] 
is non-empty and internal. By the overflow principle $G_{\epsilon,x}$ contains
an infinite $T_{\epsilon,x}$, and by polysaturation the set
\[
G_{x} = \bigcap_{\epsilon > 0} [T'_\epsilon, T_{\epsilon,x}]
\]
is non-empty, i.e. we can choose a $T_{-,x} \in G_{x}$. $T_{-,x}$ is infinite 
since $T_{-,x}  > T'_\epsilon \geq 1 / \epsilon$ for all $\epsilon > 0$.
Moreover, if $t$ is infinite and  $t \leq T_{-,x}$ we obtain $(\forall \epsilon>0) t \in 
[T'_\epsilon, T_{\epsilon,x}]$, and thus $^o y = W^- \, ^ox = \, ^o(Y_t \,x)$.
$\diamondsuit$ \\*[.3cm]
{\bf Theorem 5:}  Assume that $^oH$ contains a standard Hilbert space ${\cal H}$,
then there exist infinite hyperreals
$T_\pm$ for which $W^\pm \, ^ox = \, ^o(Y_t \, x)$ holds if $t$ is infinite,
$0 < (\mp t) \leq T_\pm$, $x \in L^\pm$, and $^ox \in {\cal H}$.\\[.15cm]
{\em Proof}: Again, we carry out only the proof for $L^-$, since the proof for $L^+$ is 
analogous. Let $x \in L^-$ and let $T_{-,x}$ be the infinite hyperreal determined by lemma 3.
Since $Y_t$ is finitely bounded for each $t \in \, ^\star \R$ we obtain 
$^o(Y_t \, x) = \, ^o(Y_t \, y)$ for all $y \approx x$. Thus, if $t$ is infinite and 
$0 < t < T_{-,x}$ then $W^- \, ^oy = \, ^o(Y_t\,y)$ for all $y \approx x$.
For $z \in {\cal H} \cap \, ^oL^-$ we choose a 
representative $x \in L^-$ for which $^ox = z$ holds, and set $T'_{-,z} = T_{-,x}$.
For $y \in L^-$ we obtain then $W^- z = \, ^o(Y_t \, y)$ if $t$ is infinite, 
$0 < t < T'_{-,z}$, and $^oy = z$. Let $G_{n,z} = [n, T'_{-,z}]$, then there exists 
an infinite  $T_- \in \bigcap_{n \in \N, z \in {\cal H}} G_{n,z}$ by polysaturation.
Since $T_- \leq T'_{-,z}$ for all $z \in {\cal H}$ we obtain the assertion.
$\diamondsuit$ \\*[.3cm]
We note that theorem 5 is valid also if we replace ${\cal H}$ by any set of standard elements.
The crucial point is that we use the properties of polysaturation in our proof, which limits
the size of subsets of $L^\pm$ for which common hyperreals $T_\pm$ can be determined.
However, if we accept that we observe only a standard set of states in a 
scattering experiment, then theorem 5 yields an interesting result.
Motivated by theorem 5, we may identify three phases of a scattering experiment:
The preparation of the system is done in the remote past at infinite times $t$ for
which $- T_+ \leq t < 0$ holds. The scattering happens at finite times $t$,
and the detection takes place in the far future at times $t$ for which 
$0 < t \leq T_-$ holds. $-T_+$ and $T_-$ may be interpreted as starting time
and as finishing time of the experiment, respectively. 

Moreover, we naturally obtain a separation of time scales within our nonstandard model: 
We observe the interacting system on a
large time scale that is defined by the infinite time interval $[-T_+, T_-]$, whereas the 
interaction takes place on a small time scale that is defined by finite times, $t \in
\fin(^\star\R)$. The separation of time scales, which cannot be done in standard
models in the same manner, is a nice example of the strength of NSA. We note however
that our result does not state that the preparation and the detection take place on the
same time scale since the fraction $T_- / T_+$ is not necessarily finite.

\subsection{Time-independent scattering theory}
\label{TIST}

We discuss time-independent scattering theory in the following on the basis of our results for
time-dependent scattering theory. In particular, we assume in this section the 
setting of theorem 5 and that $H$ is hyperfinite-dimensional. 
As outlined in the discussion of theorem 5, we may assume especially two infinite
hyperreals $T_\pm$ that mark the starting and the finishing times of a scattering experiment.
We introduce first a set of infinitesimals that is closely related to $T_\pm$.
Let $T = \min\{T_-, T_+\}$, and let 
$\Gamma = \{ t \approx 0: \, t > 0 \,\, \mbox{and} \,\, tT \, \mbox{infinite} \}$.
$\Gamma$ is non-empty since, for example, 
$\{ t/ \sqrt{T} : \, ^ot > 0 \, \mbox{and} \, t \in \fin(^\star \R_+) \} \subset  \Gamma$. 
We note that $\Gamma$ is a set of infinitesimals that depends only on $T_\pm$, and that
is thus fundamentally related to the scattering system.
We assume for the rest of the section that $\epsilon \in \Gamma$, i.e. 
$\epsilon \approx 0$, $\epsilon > 0$, and $\epsilon t$ is infinite for 
$t \geq \min\{T_-, T_+\}$. Moreover, we use in the following transferred 
integration theory.
\\[.3cm]
{\bf Lemma 4:} Let $Y^\pm = Y_{\mp T_\pm / 2}$, then
\begin{equation}
Y^\pm u \approx \int\limits_0^{^\star \infty} ds \, \epsilon e^{-\epsilon s} Y_{\mp s} u
\end{equation}
if $u \in L^\pm$ and $^ou \in {\cal H}$.
\\[.15cm]
{\em Proof}: 
Let $0< \, ^o\alpha < 1$, let $T_1 = 1 / \sqrt{\epsilon}$, and 
let $T_2 = \alpha \min\{T_+,T_-\}$. 
Since $\| Y_s \| =1$ for all $s \in \, ^\star \R$ we obtain
\begin{eqnarray*}
\left\| \int\limits_0^{^\star T_1} ds \, \epsilon e^{-\epsilon s}
Y_{\mp s} \right\| & \leq & \int\limits_0^{^\star \sqrt{\epsilon}} ds \,
e^{-s} = 1 - e^{-\sqrt{\epsilon}} \approx 0 \, ,\\
\left\| \, \int\limits_{T_2}^{^\star \infty} ds \, \epsilon e^{-\epsilon s}
Y_{\mp s} \right\| & \leq & e^{- \epsilon T_2} \approx 0 \, , \\
\int\limits_{T_1}^{^\star T_2} ds \, \epsilon e^{-\epsilon s}
Y_{\mp s} & \approx &  
\int\limits_0^{^\star \infty} ds \, \epsilon e^{-\epsilon s} Y_{\mp s}. 
\end{eqnarray*}
Assume $u \in L^\pm$ and $^ou \in {\cal H}$.
By theorem 5, $Y_{\mp s} \, u \approx Y^\pm u$ for $s \in [T_1,T_2]$, and thus
\[
\int\limits_{T_1}^{^\star T_2} ds \, \epsilon e^{-\epsilon s} Y_{\mp s} u
\approx ( e^{-\sqrt{\epsilon}} - e^{- \epsilon T_2} ) Y^\pm u \approx Y^\pm u
\quad \diamondsuit
\]
Lemma 4 enables us to deduce an explicit formula for
the operators $W^\pm = \, ^oY^\pm$. Let $\{ \lambda_j,x_j \}_{j=1}^h$ 
be an eigensystem of the free Hamiltonian $B_0$.
If $u \in L^\pm$ and $^ou \in {\cal H}$ we obtain
\begin{eqnarray*}
Y^\pm u & \approx & \int\limits_0^{^\star \infty} dt \, \epsilon e^{-\epsilon t} Y_{\mp t} u \\
& = & u \mp i \int\limits_0^{^\star \infty} dt \, e^{-\epsilon t} e^{\mp iBt} V e^{\pm iB_0t} u 
\quad (V = B - B_0) \\
& = & u \mp i \sum_{j=1}^h u_j 
\int\limits_0^{^\star \infty} dt \, e^{-\epsilon t} e^{\mp i(B - \lambda_j)t} V \, x_j \quad
(u_j = \langle x_j, u \rangle) \\
& = & u \pm i \sum_{j=1}^h u_j \frac{1}{ \mp i(B - \lambda_j) -\epsilon } V \, x_j \\
& = & \sum_{j=1}^h u_j \left(1 + \frac{1}{\lambda_j - B \pm i \epsilon } V \right) x_j \, .
\end{eqnarray*}
{\bf Theorem 6:} 
\begin{equation}
W^\pm \, ^ou = \, ^o(Y^\pm u) = \, 
^o\sum_{j=1}^h u_j \left(1 + \frac{1}{\lambda_j - B \pm i \epsilon } V \right) x_j 
\end{equation}
if $u \in L^\pm$ and $^ou \in {\cal H}$.
\\[.3cm]
Let $[Y^+, B_0] = Y^+ B_0 - B_0 Y^+$, and let $\T_t = e^{iB_0t} [Y^+, B_0]
e^{-iB_0t}$. We consider now the 'S-Matrix', $S = (Y^-)^\dagger Y^+$,
for which $\langle W^- \, ^ou, W^+ \, ^ov \rangle = \, ^o\langle u, S v
\rangle$ holds if $u \in L^-$, $v \in L^+$, and $^ou, \, ^ov \in {\cal H}$.\\[.3cm]
{\bf Lemma 5:} 
\begin{equation}
\langle u, (S-1) v \rangle \approx -i \int\limits_{- ^\star\infty}^{^\star\infty}
dt \, e^{- \epsilon |t|} \langle u, \T_t v \rangle
\end{equation}
if $u \in L^- \cap L^+$, $v \in L^+$, and $^ou, \, ^ov \in {\cal H}$.
\\[.15cm]
{\em Proof}: 
If $u \in L^- \cap L^+ $, $v \in L^+$, and $^ou, \, ^ov \in {\cal H}$ we obtain by lemma 4,
\[
\langle u, (S-1) v \rangle  = \langle u, (Y^- - Y^+)^\dagger \, Y^+ v \rangle  
\approx \int\limits_0^{^\star\infty} \epsilon e^{-\epsilon s} 
\langle u, (Y_t - Y_{-t})^\dagger \, Y^+ v \rangle .
\] 
Let $0< \, ^o\alpha < 1/2$, let $T_1 = 1 / \sqrt{\epsilon}$, and let $T_2 = \alpha 
\min\{T_+, T_- \}$, then
\[
\int\limits_0^{^\star\infty} dt \, \epsilon e^{-\epsilon t} 
\langle u, (Y_t - Y_{-t})^\dagger \, Y^+ v \rangle  \approx
\int\limits_{T_1}^{^\star T_2} dt \, \epsilon e^{-\epsilon t} 
\langle u, (Y_t - Y_{-t})^\dagger \, Y^+ v \rangle .
\]
By theorem 5, we obtain for $t \in [T_1, T_2]$
\begin{eqnarray*}
(Y_t)^\dagger \, Y^+ v & \approx & e^{iB_0t} e^{-iBt} e^{-iB(T_+/2 -t)} e^{iB_0(T_+/2 -t)} v \\
& = & e^{iB_0t} e^{-iB T_+/2} e^{iB_0 T_+/2} e^{- iB_0t} v \\
& = & e^{iB_0t} Y^+ e^{- iB_0t} v 
\end{eqnarray*}
Analogously,
\begin{eqnarray*}
(Y_{-t})^\dagger \, Y^+ v & \approx & e^{-iB_0t} e^{iBt} 
e^{-iB(T_+/2 + t)} e^{iB_0(T_+/2 + t)} v \\
& = & e^{-iB_0t} e^{-iB T_+/2} e^{iB_0 T_+/2} e^{iB_0t} v \\
& = & e^{-iB_0t} Y^+ e^{iB_0t} v 
\end{eqnarray*}
Let $Z_t = e^{iB_0t} Y^+ e^{- iB_0t}$, then
\begin{eqnarray*}
\int\limits_{T_1}^{^\star T_2} dt \, \epsilon e^{-\epsilon t} 
\langle u, (Y_t - Y_{-t})^\dagger \, Y^+ v \rangle & \approx &
\int\limits_{T_1}^{^\star T_2} dt \, \epsilon e^{-\epsilon t} 
\langle u, (Z_t - Z_{-t}) v \rangle \\
& \approx & \int\limits_0^{^\star\infty} dt \, \epsilon e^{-\epsilon t} 
\langle u, (Z_t - Z_{-t}) v \rangle .
\end{eqnarray*}
We recognize that $^\star(d/dt) Z_t = -i \T_t$, and obtain by partial integration
\begin{eqnarray*}
\int\limits_0^{^\star\infty} dt \, \epsilon e^{-\epsilon t} 
\langle u, (Z_t - Z_{-t}) v \rangle & = & -i
\int\limits_0^{^\star\infty} dt \, e^{-\epsilon t} 
\langle u, (\T_t + \T_{-t}) v \rangle \\
& = & -i
\int\limits_{-^\star\infty}^{^\star\infty} dt \, e^{-\epsilon |t|} 
\langle u, \T_t v \rangle . \quad \diamondsuit
\end{eqnarray*}
Let $T_{j,k} = \langle x_j, [Y^+, B_o] x_k \rangle$ (the ''T-matrix''). For
$u \in L^- \cap L^+ $, $v \in L^+$, and $^ou, \, ^ov \in {\cal H}$ we obtain
\begin{eqnarray*}
\langle u, (S-1) v \rangle & \approx & -i \int\limits_{- ^\star\infty}^{^\star\infty}
dt \, e^{- \epsilon |t|} \langle u, \T_t v \rangle \\
& = & -i \sum_{j,k =1}^h u_j^\ast v_k \,T_{j,k} \int\limits_{- ^\star\infty}^{^\star\infty}
dt \, e^{- \epsilon |t|} e^{i(\lambda_j - \lambda_k)} \\
& & \qquad (u_j^\ast = \langle u, x_j \rangle, \, v_k = \langle x_k, v \rangle) \\
& = & -i \sum_{j,k =1}^h u_j^\ast v_k \,T_{j,k} 
\left( \frac{1}{i(\lambda_j - \lambda_k) + \epsilon }
- \frac{1}{i(\lambda_j - \lambda_k) - \epsilon } \right) \\
& = & -i \sum_{j,k =1}^h u_j^\ast v_k \,T_{j,k} \,
\frac{-2 \epsilon}{-(\lambda_j - \lambda_k)^2 - \epsilon^2} \\
& = & - 2 \pi i \sum_{j,k =1}^h u_j^\ast v_k \,\delta_\epsilon(\lambda_j - \lambda_k) T_{j,k}.
\end{eqnarray*}
We note that since $\epsilon$ is a positive infinitesimal the function 
\begin{equation}
\delta_\epsilon(x) = \frac{1}{\pi} \, \frac{\epsilon}{x^2 + \epsilon^2} 
\end{equation}
behaves like Dirac's delta distribution \cite{far75:177}. 
\\[.3cm]
{\bf Theorem 7} 
\begin{equation}
\langle ^ou, \, ^oS \,^ov \rangle = \,
^o\sum_{j,k =1}^h \Big( \delta_{j,k} - 2 \pi i \delta_\epsilon(\lambda_j - \lambda_k) T_{j,k}
\Big) u_j^\ast v_k
\end{equation}
if $u \in L^- \cap L^+ $, $v \in L^+$, and $^ou, \, ^ov \in {\cal H}$.
\\[.15cm]

\section{Summary and conclusions}
\label{Conc}

We have formulated in this article an approach to nonstandard quantum mechanics, 
i.e. we have introduced a mathematical framework that is based on NSA, and that
can be appropriately applied to quantum mechanics. The principal step is the embedding 
of a dense subset of a standard complex Hilbert space ${\cal H}$ 
into a hyperfinite-dimensional space $H$. 
We have focused then on self-adjoint operators on ${\cal H}$, and have constructed appropriate 
extensions to self-adjoint hyperfinite-rank operators on $H$. To obtain a sensible 
interpretation of expectation values we have introduced further nonstandard hulls. 
We have defined the nonstandard hull $^oH$ of the hyperfinite-dimensional space $H$, which is
a complex Hilbert space, and the nonstandard hulls of hyperfinite-rank operators on $H$. 
The idea is to perform calculations in $H$, and to interpret expectation values in the 
nonstandard hull $^oH$. To this end, we have developed the function calculus with respect to 
Loeb-measurable functions by introducing projection-valued Loeb measures. Using 
projection-valued Loeb measures we have proved then a nonstandard spectral theorem.
With the help of the spectral theorem we have shown how our nonstandard framework
extends the standard framework used for quantum mechanics. As an example, we have 
applied our general results to Schr\"odinger representations of quantum mechanics.

Beside the mathematical results, which are rather interesting for their own, two important 
advantages are obtained by this approach. First, we can use nonstandard objects like 
infinitesimals, infinite numbers, or functions that behave like delta distributions in our 
framework. The availability of these objects is a general advantage of NSA as compared to 
standard mathematics. Second, since we perform calculations in a hyperfinite-dimensional 
space we can apply the transferred rules of linear algebra. In particular, 
self-adjoint hyperfinite-rank operators have complete sets of eigenvalues and eigenvectors. 
If we use this result to model the Hamiltonian of 
a quantum system we treat bound states and continuum states on the same footing. We note 
that this is also a feature of super-Hilbert space formalisms \cite{ant_boh98,boh_boh98}, 
but the present approach seems to be more convenient for physical applications.

Moreover, we have applied the approach to non-relativistic scattering theory. First we have 
extended standard time-dependent scattering theory to our framework. If we observe a standard 
set of states in a scattering experiment, as stated in theorem 5, then the extension yields 
two fundamental times of a scattering experiment that can be interpreted as the 
starting time, $-T_+$, and the finishing time, $T_-$, of the experiment. 
These times occur as infinite hyperreals, and are not available in standard
theories. Moreover, this result yields a natural separation of time scales: We observe the 
interacting system on a large time scale that is defined by the infinite time interval 
$[-T_+, T_-]$, whereas the interaction takes place on a small time scale that is 
defined by finite times, $t \in \fin(^\star\R)$. In particular, the preparation of the system 
is done in the remote past at infinite times $t$ for which $-T_+ \leq t < 0$ holds, and the 
detection takes place in the far future at infinite times $t$ for which $0 < t \leq T_-$ holds. 
We note however that this result does not state that the 
preparation and the detection processes take place on the
same time scale since the fraction $T_- / T_+$ is not necessarily finite. 
Furthermore, we have applied our results to time-independent scattering theory. For
time-independent scattering theory we have shown how concrete calculations work in the 
framework, and we have derived explicit formulas for the M\o ller wave operators and for 
the S-Matrix. These formulas, which are well-known from standard physical text books, 
are proved rigorously.

From my point of view, the application of nonstandard methods to non-relativistic scattering 
theory is rather fruitful. In particular, this example demonstrates the main 
advantages of the formalism mentioned above. 

\appendix

\section*{Appendix}

The linear algebra in finite-dimensional subspaces of ${\cal H}$
is well-known from standard text books. We need the results however to obtain the
linear algebra in hyperfinite-dimensional subspaces of $^\star {\cal H}$ in the following.
For this purpose, we formulate the standard results in a way that enables us to easily 
apply the transfer principle. However, although a presentation of nonstandard linear 
algebra should be available elsewhere I was not able to find an appropriate reference 
in the literature.

\subsection*{Linear algebra in finite-dimensional subspaces}

Let ${\cal F}$ be the set of finite-dimensional subspaces of ${\cal H}$. For $F \in {\cal F}$
let $\ONB (F)$ be the set of orthonormal bases in $F$. The dimension $d$ of $F$ is 
the unique number of elements of each $B \in \ONB (F)$, i.e.
\begin{equation}
(\exists d \in \N) \, (\forall B \in \ONB (F) ) \quad |B| = d \, .
\end{equation}
We use here $\ONB$ as a function on ${\cal F}$. Let $\delta : 
{\cal H} \times {\cal H} \to \{ 0,1 \}$ be the function defined by 
\[
\delta(x,y) = 
\left\{ \begin{array}{c@{\quad,\quad}c} 1 & x = y \\
0 & \mbox{else} \end{array} \right. \, .
\]
The orthonormality of the bases can be expressed then as
\[
(\forall B \in \ONB (F) ) \,(\forall x,y \in B ) \quad
\langle x,y \rangle = \delta (x,y) \, ,
\]
and basis-set expansions are given by
\[
(\forall B \in \ONB (F) ) \,(\forall x \in F) 
\quad x = \sum_{y \in B} \langle y,x \rangle \, y \, .
\]
We denote the set of finite-rank operators on ${\cal H}$ by
\begin{equation}
\LF = \{ f \in {\cal H}^{\cal H} \, : \, ´
\mbox{ $f$ is linear and} \, \, R(f) \in {\cal F} \} \, . 
\end{equation}
$R(f)$ denotes the range of the function $f$. A finite-rank operator $A$ on ${\cal H}$ is 
called normal if $A^\dagger A = A A^\dagger$,
and we denote the set of normal finite-rank operators on ${\cal H}$ by
\begin{equation}
\LFN = \{ A \in \LF \, : \, A^\dagger A = A A^\dagger \} \, .
\end{equation}
For $A \in \LF$ and $B \in \ONB (R(A))$ we can express the action of $A$ on $R(A)$ as follows
\[
(\forall x \in R(A)) \quad A x = \sum_{ y \in B}  \langle y, A x \rangle \, y =
\sum_{ y, y' \in B}  \langle y, A y' \rangle \, \langle y',x \rangle \, y \, .
\]
If $A$ is normal, then $(\langle y, A y' \rangle)_{ y, y' \in B}$ is a normal matrix.
Hence, there exists an eigensystem that can be used to express the action of $A$ on $R(A)$,
and we obtain
\begin{eqnarray}
\nonumber
(\forall A \in \LFN) \, (\exists B \in \ONB (R(A)) ) \, (\exists f \in \C^{\cal H} ) \,
(\forall x \in R(A))  & &\\
A x = \sum_{ y \in B} f(y) \langle y, x \rangle \, y \, . & & 
\end{eqnarray}
The set $\{ f(y) \}_{ y \in B}$ is the set of eigenvalues of the restriction of $A$ to $R(A)$.

\subsection*{Linear algebra in hyperfinite-dimensional subspaces}

We assume in the sequel a polysaturated extension of the basic superstructure.
Then, $^\star {\cal F}$ is the set of hyperfinite-dimensional subspaces of $^\star {\cal H}$.
The transferred relation $^\star \ONB$ acts on $^\star {\cal F}$, and by the transfer 
principle we obtain for $F \in \, ^\star {\cal F}$
\begin{equation}
(\exists d \in \, ^\star \N) \, (\forall B \in \, ^\star 
\ONB (F) ) \quad |B| = d \, .
\end{equation}
We note that the dimension $d$ of $F$ is now a hypernatural number, which may be infinite.
The orthonormality of the nonstandard bases is expressed by
\[
(\forall B \in \, ^\star \ONB (F) ) \, (\forall x,y \in B ) \quad 
^\star \langle x,y \rangle = \, ^\star \delta (x,y) \, ,
\]
and the nonstandard basis-set expansions are given by
\[
(\forall B \in \, ^\star \ONB (F) ) \, (\forall x \in F) \quad x = \sum_{y \in B} 
\, ^\star \langle y,x \rangle \, y \, .
\]
Note that $^\star \langle \cdot , \cdot \rangle$ is the transferred scalar product, and
that $\sum_{y \in B} $ may be a hyperfinite sum.
The internal hyperfinite-rank operators on $^\star {\cal H}$ are given by the set
\begin{equation}
^\star\LF = \{ f \in \, ^\star ({\cal H}^{\cal H}) \, : \, 
\mbox{ $f$ is linear and} \,\,  R(f) \in \, ^\star{\cal F} \} \, . 
\end{equation}
A normal linear operator $A$ on $^\star {\cal H}$ fulfils the condition
$A^\dagger A = A A^\dagger$,
and the set of normal hyperfinite-rank operators on $^\star {\cal H}$ is 
\begin{equation}
^\star \LFN = \{ A \in \, ^\star \LF \, : \, A^\dagger A = A A^\dagger \} \, .
\end{equation}
For each $A \in \, ^\star \LFN$ there exists an eigensystem, and we obtain
\begin{eqnarray}
\nonumber
(\forall A \in \, ^\star \LFN) \, (\exists B \in \, ^\star \ONB (R(A)) ) \,
(\exists f \in \, ^\star (\C^{\cal H}) ) \,
(\forall x \in R(A)) & & \\
A x = \sum_{ y \in B} f(y) \,\,  ^\star \langle y, x \rangle \, y \, . & &
\end{eqnarray}
Since each $A \in \, ^\star \LFN$ is internal, the range of $A$, $R(A)$, is internal,
$^\star \ONB (R(A))$ is internal, and each $B \in \, ^\star \ONB (R(A))$ is internal.
Moreover, $f \in \, ^\star (\C^{\cal H})$ is internal, and the set of 
eigenvalues, $f(B)$, is thus internal too.

If we consider a standard eigensystem $E$ then it is convenient to enumerate the elements,
$E = \{ (\lambda_1, x_1) , ... , (\lambda_d, x_d)\}$, and $d \in \N$ is the dimension.
In the same manner we can denote a nonstandard eigensystem, assuming $d \in \,^\star \N$.

\end{document}